\documentclass[
aps,
prl,           
twocolumn,
longbibliography,
nobibnotes,
floatfix,      
]{revtex4-2}

\usepackage{graphicx}
\usepackage{amsmath}
\usepackage{hyperref}
\usepackage{physics}
\usepackage{float}

\hypersetup{
	colorlinks=true,
	linkcolor=blue,
	citecolor=blue,
	urlcolor=blue
}

\begin{document}

\title{Always-on, highly efficient microwave photon detector based on a superconducting artificial molecule}

\author{Vyom Kulkarni}
\email{vyom@chalmers.se}
\author{Mohammed Ali Aamir}
\author{Simon Sundelin}
\author{Simone Gasparinetti}
\email{simoneg@chalmers.se}
\affiliation{Department of Microtechnology and Nanoscience, Chalmers University of Technology, 412 96 Gothenburg, Sweden}

\date{\today}

\begin{abstract}
Efficient detection of single microwave photons is a key capability for emerging quantum technologies. Yet, it remains far less developed than its optical domain counterpart. Realizing detectors that simultaneously achieve high efficiency, low dark counts, and continuous operation has proved challenging. Existing detectors operate cyclically, forcing a trade-off between efficiency and duty cycle. Here, we demonstrate a continuously operated microwave single-photon detector based on a superconducting artificial molecule. In our scheme, an incoming photon is captured by a bright state of the molecule and then transferred to a long-lived dark state via a driven-dissipative process. Photon ``clicks'' are revealed as quantum jumps in the continuously monitored dark state. We observe a cyclic detection efficiency of $0.73$, and a continuous detection efficiency of $0.47$ over a $5\,\mathrm{MHz}$ instantaneous bandwidth, with a $1\,\mu\mathrm{s}$ temporal resolution and a $15\,\mu\mathrm{s}$ dead time. By overcoming the trade-off between efficiency and duty cycle, this approach establishes continuous microwave photon detection for quantum sensing, quantum thermodynamics, and fundamental physics.
\end{abstract}

\maketitle

\section{I. INTRODUCTION}
Single-photon detectors at optical frequencies~\cite{hadfield_single-photon_2009} have enabled fundamental tests of quantum physics~\cite{weihs1998, hong1987, grangier1986} and the development of quantum technologies~\cite{magde1974, moerner1989, liao2018a}. Extending this capability to the microwave frequency domain would enable novel approaches to distributed quantum computing~\cite{narla2016}, qubit readout~\cite{govia2014}, and especially quantum sensing~\cite{wang2023,albertinale2021,keranen2025, braggio2025, dixit2021}. However, detecting single microwave photons has proved difficult due to the low energies involved. Furthermore, in many sensing applications, the arrival time of the photons to be detected is stochastic, so the detector must be operated continuously. Despite these challenges, early microwave single-photon detectors have been applied to the sensing of spin fluorescence~\cite{wang2023, albertinale2021}, measurement of correlations in propagating microwaves~\cite{keranen2025}, and the monitoring of cavity-haloscope emission for axion search experiments~\cite{braggio2025, dixit2021}.

Current approaches to microwave photon detection rely on nanobolometers~\cite{lee2020e, chang2025, kokkoniemi2020}, Josephson junctions~\cite{chen2011a, albert2024}, semiconductor quantum dots~\cite{oppliger2025, haldar_continuous_2024, haldar2024b, khan2021}, and superconducting qubits~\cite{inomata2016, kono2018, besse2018, lescanne2020a, balembois2024, pallegoix2025}. In nanobolometers, photon absorption is inferred from a temperature rise in the absorber, with a demonstrated energy resolution of the order of tens of GHz. Josephson-junction-based detectors and semiconductor quantum dots convert photon absorption into a change in voltage or current but currently lack sufficiently fast readout to enable single-shot operation. As a result, although these detectors can operate continuously, they have not yet demonstrated the ability to resolve individual events.

By contrast, in superconducting qubit-based detectors, the information on the presence of the photon is mapped onto the state of a qubit, which can be read out in a single shot with high fidelity~\cite{jeffrey2014, walter2017}. Nondestructive schemes based on a Ramsey-type sequence~\cite{kono2018, besse2018} as well as schemes based on direct photon absorption by the qubit~\cite{inomata2016,lescanne2020a, balembois2024, pallegoix2025} have been demonstrated. Crucially, both of these schemes have only been demonstrated in cyclic operation, with detection cycles consisting of a photon acquisition window, strong projective measurement of the qubit, and qubit reset~\cite{balembois2024}. While photons arriving during the acquisition window are detected, photons arriving during measurement or reset are missed. In principle, these detectors could be operated continuously. However, this presents some challenges of its own. For instance, a strong pump, when active continuously, may cause heating of the photodetector and its environment~\cite{lescanne2020a}. In addition, measurement backaction from continuous monitoring of the qubit may affect the detection efficiency~\cite{clerk2010a}.

Here, we demonstrate a novel type of microwave photon detector in a superconducting circuit. We achieve continuous detection with an efficiency of $0.47$ and a dark count rate of $1.3\,\mathrm{kHz}$. We also observe a trade-off between measurement backaction and detection efficiency.

\begin{figure*}[!t]
    \centering
    \includegraphics[width=\textwidth]{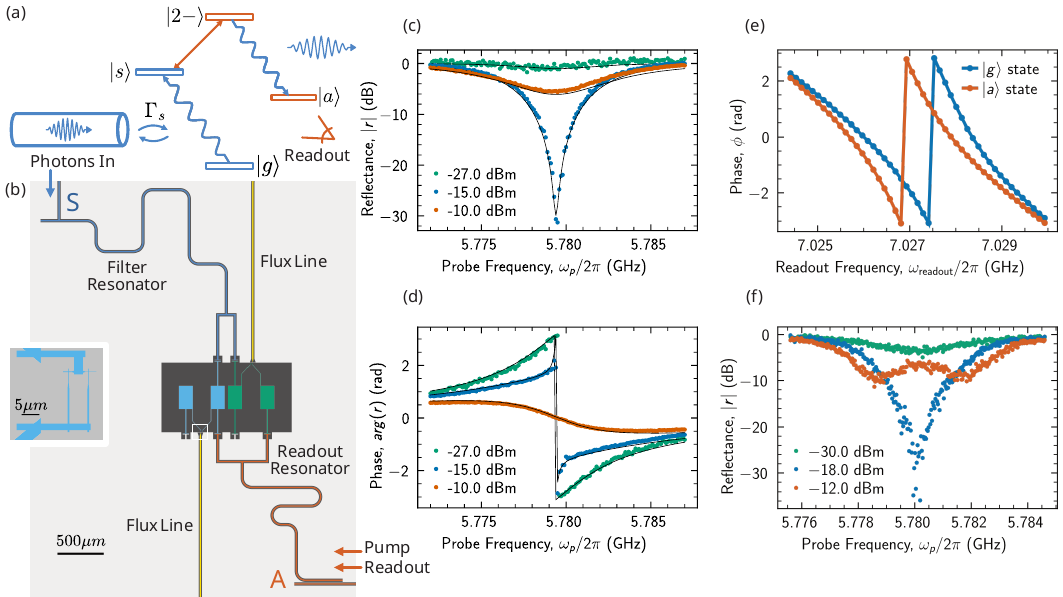}
    \caption{
    \textbf{a.} Operating principle of the detector. Incident photons are absorbed by the symmetric mode. The pump coherently excites the system to the $\ket{2-}$ state, which subsequently relaxes to the metastable state $\ket{a}$. The population of $\ket{a}$ is continuously monitored using dispersive readout.
    \textbf{b.} Device layout. The circuit consists of two coupled flux-tunable transmon qubits, shown in light blue and green, which are tuned into resonance using flux lines, shown in yellow. The inset displays the superconducting quantum interference device located in close proximity to the flux line. The device has two radio-frequency ports: waveguide~S, which routes incident photons to the detector, and waveguide~A, which carries the pump and readout tones.
    \textbf{c,d.} Symmetric mode spectroscopy. Magnitude and phase response of the reflection coefficient $r$ measured across the symmetric mode frequency $\omega_s/2 \pi$ for three different probe powers. Solid curves indicate a global fit to the data, as described in \hyperref[app:sm_fit]{Appendix~B}.
    \textbf{e.} Dispersive readout of $\ket{a}$. The phase response $\phi$ of the readout resonator is shown as a function of readout frequency for the artificial molecule prepared in either $\ket{g}$ or $\ket{a}$.
    \textbf{f.} Photon-transfer mechanism. The reflection coefficient $r$ is measured as a function of probe frequency for three different pump powers. The pump is tuned in resonance with the $\ket{s}\leftrightarrow\ket{2-}$ transition, as shown in \hyperref[app:pump]{Appendix~D}.
    }
    \label{fig:fig1}
\end{figure*}

\section{II. RESULTS}
\subsection{A. DEVICE DESIGN AND CHARACTERIZATION}
Our detector uses a driven-dissipative absorption process in a four-level system (\hyperref[fig:fig1]{Fig.~\ref*{fig:fig1}a}). An incoming photon resonant with the frequency of mode $\ket{s}$ is transferred to state $\ket{a}$ via a combination of coherent pump and relaxation mediated by a higher-order state $\ket{2-}$. We implement this four-level system in a superconducting circuit that comprises two nominally identical, frequency-tunable transmon qubits~\cite{koch2007a} coupled together (\hyperref[fig:fig1]{Fig.~\ref*{fig:fig1}b}). The circuit is connected to two waveguides: waveguide~S, through which incident photons are routed to the detector, and waveguide~A, which carries the pump and readout tones. When the two qubits are brought into resonance, the Hamiltonian of the combined system commutes with the qubit permutation operator, so that all eigenstates are either symmetric or anti-symmetric with respect to spatial exchange of the two qubits. In the single-excitation manifold specifically, the eigenstates of the molecule are the anti-symmetric and symmetric combinations of the qubit states $\ket{10}$ and $\ket{01}$, which we denote $\ket{a} = (\ket{10} - \ket{01})/\sqrt{2}$ and $\ket{s} = (\ket{10} + \ket{01})/\sqrt{2}$, respectively, with measured transition frequencies $\omega_{a}/2\pi = 5.326\,\mathrm{GHz}$ and $\omega_{s}/2\pi = 5.780\,\mathrm{GHz}$. Furthermore, in the double-excitation manifold, our protocol directly involves the state $\ket{2-} = (\ket{20} - \ket{02})/\sqrt{2}$, which is of the anti-symmetric type at a transition frequency $\omega_{s,2-}/2\pi = 4.946\,\mathrm{GHz}$ (\hyperref[app:pump]{Appendix~D}).

To engineer the desired decay rates, we exploit the symmetry of the device. By coupling additional circuitry to multiple spatial points of the two qubits and using interference effects, we engineer selection rules for transition rates between states with the same, or opposite, symmetry. This arrangement, referred to as \textit{symmetry-selective couplings}, was previously demonstrated in Ref.~\cite{aamir2022}. Here, we combine this effect with the Purcell effect provided by additional resonators~\cite{jeffrey2014}. The molecule is dispersively coupled to a readout resonator, with a frequency $\omega_{r}/2\pi = 7.027\,\mathrm{GHz}$ and decay rate $\kappa_{r}/2\pi = 1.8\,\mathrm{MHz}$, which is in turn connected to waveguide~A. This resonator protects the anti-symmetric mode from symmetry-enhanced decay into waveguide~A by the Purcell effect, extending the lifetime of the anti-symmetric mode. Furthermore, the symmetric mode is also protected from decaying into waveguide~A. The filter resonator at frequency $\omega_{f}/2\pi = 5.652\,\mathrm{GHz}$ has a broad linewidth, $\kappa_{f}/2\pi = 34\,\mathrm{MHz}$, and is coupled to waveguide~S. It is engineered to enhance the linewidth of the symmetric mode while suppressing the decay of the anti-symmetric mode. Consequently, the anti-symmetric mode is protected from decay into waveguide~S by two independent mechanisms: the intrinsic symmetry-dependent selectivity and the Purcell effect provided by the filter resonator. This results in a long-lived anti-symmetric mode and a large linewidth symmetric mode that is overcoupled to waveguide~S.

We measure the device in a dilution refrigerator at $8\,\mathrm{mK}$. We connect each waveguide to a reflection-measurement chain through circulators. The readout chain for waveguide~A includes a traveling-wave parametric amplifier as its first amplifier, and has a measurement efficiency $\eta = 0.11$ (\hyperref[app:measurement_rate]{Appendix~G}).

We characterize the symmetric mode by studying the power-dependent reflection coefficient $r$ measured from waveguide~S with a Vector Network Analyzer (VNA) (\hyperref[fig:fig1]{Fig.~\ref*{fig:fig1}c,d}). At low power (green curve), the mode is only weakly excited and coherent scattering dominates. $r$ shows a weak response in magnitude, but a full $2\pi$ phase flip, implying that the mode is overcoupled to the waveguide. Increasing the probe power produces more incoherent scattering as the mode becomes increasingly excited, resulting in a dip in the amplitude and a suppression of the phase flip. At high enough powers, the mode saturates and the dip becomes less pronounced. By globally fitting a model based on the Lindblad master equation and input-output theory to the measured $r$ and its power dependence, we extract the coupling rate $\Gamma_{s}/2\pi = 7.26\,\mathrm{MHz}$ as well as the total attenuation of the input line to waveguide~S (\hyperref[app:sm_fit]{Appendix~B}). Importantly, this non-linear response of the symmetric mode allows for calibration of the power reaching the device.

We characterize the anti-symmetric mode with pulsed measurements using a microwave transceiver. We deterministically excite the anti-symmetric mode with a $\pi$ pulse and measure its state by performing dispersive readout via the readout resonator~\cite{blais2004}. By measuring the resonator response as a function of frequency after preparing the artificial molecule in both state $\ket{g}$ and $\ket{a}$, we extract a dispersive shift $2\chi_a/2\pi = 0.87\,\mathrm{MHz}$ (\hyperref[fig:fig1]{Fig.~\ref*{fig:fig1}e}).

To demonstrate efficient absorption of microwave photons by the device, we measure reflection from waveguide~S at low power while simultaneously applying a coherent pump tone resonant with the $\ket{s}\leftrightarrow\ket{2-}$ transition (\hyperref[fig:fig1]{Fig.~\ref*{fig:fig1}f}). At low pump powers, the spectrum resembles the low-power response in \hyperref[fig:fig1]{Fig.~\ref*{fig:fig1}c}. As the pump power increases, the dip becomes more pronounced, indicating absorption of the probe. At the optimal power, we observe a dip exceeding $25\,\mathrm{dB}$. At even higher powers, we observe the Autler-Townes splitting of the $\ket{s}\leftrightarrow\ket{2-}$ transition, which calibrates the Rabi rate of the pump (\hyperref[app:pump]{Appendix~D}).

\begin{figure}[!t]
  \centering
  \includegraphics[width=\columnwidth]{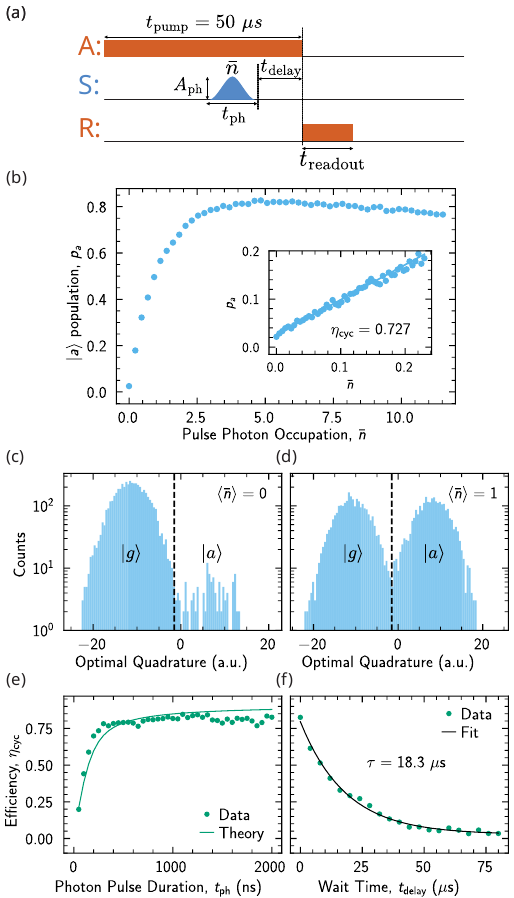}
  \caption{%
  \textbf{Cyclic photodetection.}
  \textbf{a.} Pulse sequence. A $50\,\mu\mathrm{s}$-long pump is applied to the $\ket{s}\leftrightarrow\ket{2-}$ transition. The state of $\ket{a}$ is measured at the end of the pump using a readout pulse. A coherent photon pulse with calibrated mean photon number $\bar{n}$ is routed to the detector input a time $t_\mathrm{delay}$ before the readout pulse. 
  \textbf{b.} Detection efficiency. Population $p_a$ of state $\ket{a}$ is measured using single-shot statistics as a function of mean photon number $\bar{n}$ at zero $t_\mathrm{delay}$. Inset: $p_a$ as a function of $\bar{n}$ in the range $\bar{n}\leq0.2$. The solid blue line indicates a linear fit.
  \textbf{c,d.} Representative single-shot histograms for $\bar{n}=0$ and $\bar{n}=1$.
  \textbf{e.} Cyclic detection efficiency $\eta_\mathrm{cyc}$ vs the photon pulse duration $t_\mathrm{ph}$ at zero $t_\mathrm{delay}$. Solid line shows theoretical predictions from our model, as described in \hyperref[app:theory]{Appendix~E}.
  \textbf{f.} Cyclic detection efficiency $\eta_\mathrm{cyc}$ vs delay between the photon pulse and the readout pulse $t_\mathrm{delay}$, for a pulse duration $t_\mathrm{ph}= 500\,\mathrm{ns}$.
  }
  \label{fig:fig2}
\end{figure}

\subsection{B. CYCLIC PHOTODETECTION}
To demonstrate cyclic photodetection~\cite{balembois2024}, we apply a $50\,\mu\mathrm{s}$-long pump pulse resonant with the $\ket{s}\leftrightarrow\ket{2-}$ transition. Immediately after the pump pulse, we measure the state of the anti-symmetric mode using single-shot readout~\cite{jeffrey2014, walter2017}. During the pump pulse, we irradiate the detector using a coherent microwave pulse with a calibrated mean photon number $\bar{n}$ (\hyperref[fig:fig2]{Fig.~\ref*{fig:fig2}a}). At low powers, the pulse predominantly populates the Fock states $\ket{0}$ and $\ket{1}$ of the propagating mode, thereby realizing a probabilistic single-photon source.

We first characterize the detection efficiency of the cyclic protocol. The incident photon pulses have a Gaussian envelope of total duration $t_\mathrm{ph} = 500\,\mathrm{ns}$, and arrive at the end of the detection window, immediately before the readout pulse. We vary the mean photon number $\bar{n}$ of the pulse and extract the population of the anti-symmetric mode $p_a$ from single-shot statistics (\hyperref[fig:fig2]{Fig.~\ref*{fig:fig2}b}). At low $\bar{n}$, $p_a$ increases linearly. As $\bar{n}$ is increased further, the slope decreases and $p_a$ eventually saturates at approximately $0.8$. By performing a linear fit to the data in the range $\bar{n}\leq 0.2$, we extract a cyclic single-shot detection efficiency of $\eta_\mathrm{cyc}=0.727$. At $\bar{n}\sim1$, the photon pulse begins to significantly populate higher Fock states, such that in a fraction of the shots, the detector is irradiated with more than one photon. The saturation of the population indicates that after a photon excites the anti-symmetric mode, subsequent photons are not registered. For further analysis, we study the single-shot histograms at $\bar{n}=0$ and $\bar{n}=1$ (\hyperref[fig:fig2]{Fig.~\ref*{fig:fig2}c-d}). In the absence of a photon pulse ($\bar{n}=0$), the fraction of readout shots yielding state $\ket{a}$ is $p_a = 0.025$. We estimate the dark counts as $\Gamma_\mathrm{dark}^\mathrm{cyc} = p_a/T_1 = 1.38\,\mathrm{kHz}$, where $T_1 = 18\,\mu\mathrm{s}$ is the relaxation time of state $\ket{a}$. We attribute the dark counts primarily to residual thermal photons in waveguide~S. Independent single-shot readout measurements indicate a transition error of $0.005$ for preparations in state $\ket{g}$ that are found in state $\ket{a}$ (\hyperref[app:singleshot]{Appendix~C}). These could arise from random thermal excitation of $\ket{a}$ or readout-induced transitions, implying that approximately $80\%$ of the dark counts originate from thermal photons captured from waveguide~S. The corresponding effective temperature of waveguide~S is $33\,\mathrm{mK}$, as determined from the equation~\cite{balembois2024}:
\begin{equation}
    \Gamma_\mathrm{dark}^\mathrm{cyc} = \frac{\kappa_D \eta_\mathrm{cyc}}{4}n_\mathrm{th,s}
\end{equation}
where $\kappa_D/2\pi = 5.37\,\mathrm{MHz}$ (\hyperref[app:bandwidth]{Appendix~F}) is the detection bandwidth and $n_\mathrm{th,s}$ is the residual thermal population of waveguide~S.

Next, we measure $\eta_\mathrm{cyc}$ while varying the photon pulse length $t_\mathrm{ph}$ (\hyperref[fig:fig2]{Fig.~\ref*{fig:fig2}e}). For short $t_\mathrm{ph}$, the efficiency increases rapidly and saturates at higher pulse lengths. The efficiency is limited by the imperfect spectral overlap between the pulse and the detection bandwidth. For longer pulses, the theoretical model (\hyperref[app:theory]{Appendix~E}) predicts that the efficiency asymptotically approaches unity. However, the model does not account for the finite relaxation time of the anti-symmetric mode, which explains the discrepancy between theory and experiment at longer pulse durations. For our detection bandwidth, the optimal trade-off occurs at $t_\mathrm{ph} \approx 500\,\mathrm{ns}$.

We further study the dependence of detection efficiency on the photon arrival time by varying the time interval $t_\mathrm{delay}$ between the photon pulse and the readout pulse (\hyperref[fig:fig2]{Fig.~\ref*{fig:fig2}f}). The efficiency decays exponentially as a function of $t_\mathrm{delay}$, at rate $1/T_1$. We infer from this that photons arriving early in the detection window are exponentially less likely to be detected than those arriving near its end.

\subsection{C. CONTINUOUS PHOTODETECTION}
\begin{figure}[!t]
  \centering
  \includegraphics[width=\columnwidth]{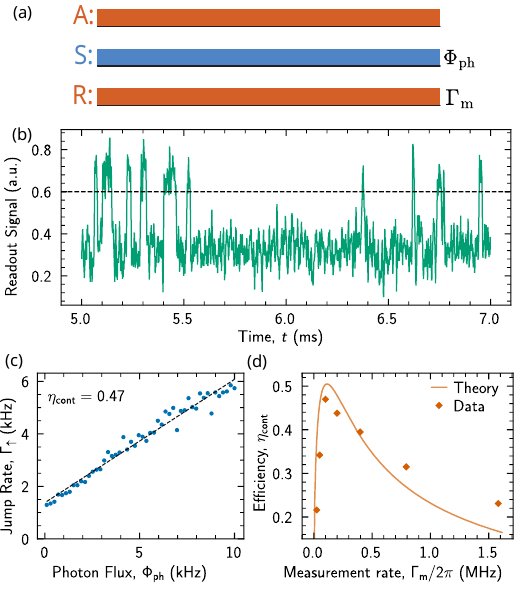}
  \caption{%
  \textbf{Continuous photodetection.}
  \textbf{a.} The pump tone is played continuously while routing a calibrated coherent photon flux $\Phi_\mathrm{ph}$ to the symmetric mode. The state of the anti-symmetric mode is measured at rate $\Gamma_\mathrm{m}$ using a continuous tone on the readout resonator.
  \textbf{b.} A typical $2\,\mathrm{ms}$-long filtered trace exhibiting 10 quantum jumps between the states $\ket{g}$ and $\ket{a}$ triggered by captured photons. The dashed black line denotes the threshold above which readout signal is counted as a jump.
  \textbf{c.} The jump rate, extracted from the time trace, is plotted against the photon flux incident on the symmetric mode. Dotted line indicates a linear fit.
  \textbf{d.} Continuous detection efficiency $\eta_\mathrm{cont}$ is plotted against the measurement rate $\Gamma_\mathrm{m}$. Solid line indicates predictions from our theoretical model using independently calibrated parameters.
  }
  \label{fig:fig3}
\end{figure}

To implement continuous photodetection, we apply a continuous pump tone resonant with the $\ket{s}\leftrightarrow\ket{2-}$ transition. Simultaneously, we measure the state of the anti-symmetric mode using a continuous readout tone. To characterize the detection, we irradiate the symmetric mode with a constant coherent photon flux $\Phi_\mathrm{ph}$ (\hyperref[fig:fig3]{Fig.~\ref*{fig:fig3}a}). Due to our engineered absorption mechanism, the incident photon flux induces quantum jumps between $\ket{g}$ and $\ket{a}$~\cite{vijay2011}. We continuously acquire the readout signal using a microwave transceiver. The time trace is integrated in real time in $1\,\mu\mathrm{s}$ bins and then stored on the measurement computer. To identify jump events, we apply a matched filter to the measurement record. We choose a five-tap kernel that rises exponentially at the relaxation rate of the anti-symmetric mode. We then count jumps by identifying peaks in the filtered record above a constant threshold. A representative measurement record with 10 such jumps is shown in \hyperref[fig:fig3]{Fig.~\ref*{fig:fig3}b}. Independently, we use the dwell time distribution of the $\ket{a}$ state to estimate the dead time of the detector, which we find to be $t_\mathrm{dead} = 15\,\mu\mathrm{s}$ (\hyperref[app:deadtime]{Appendix~H}), set by the natural relaxation of the anti-symmetric mode.

We then study the jump rate as a function of the photon flux incident on the symmetric mode $\Phi_\mathrm{ph}$ (\hyperref[fig:fig3]{Fig.~\ref*{fig:fig3}c}). We find that the jump rate increases linearly with the incident photon flux. We extract a detection efficiency $\eta_\mathrm{cont} = 0.47$ from the slope and a dark count rate $\Gamma_\mathrm{dark}^\mathrm{cont} = 1.3\,\mathrm{kHz}$ from the $y$-axis intercept. 

We next study the detection efficiency $\eta_\mathrm{cont}$ as a function of the measurement rate $\Gamma_\mathrm{m}$ (\hyperref[fig:fig3]{Fig.~\ref*{fig:fig3}d}). The measurement rate $\Gamma_\mathrm{m} = \eta \Gamma_{\phi}$ is the rate at which we extract information about the state of the anti-symmetric mode, where $\eta=0.11$ is the measurement efficiency of the readout chain and $\Gamma_{\phi}$ is the measurement-induced dephasing rate \cite{clerk2010a}. We calibrate $\eta$ and $\Gamma_{\phi}$ independently using an ac-Stark shift measurement on the anti-symmetric mode (\hyperref[app:measurement_rate]{Appendix~G}). The detection efficiency $\eta_\mathrm{cont}$ peaks at $\Gamma_\mathrm{m}/2\pi = 0.10\,\mathrm{MHz}$. Note that in this measurement, for the two lowest measurement rates, we use a matched filter kernel with 25 taps to improve the signal-to-noise ratio. 

At low measurement rates, the detection efficiency is limited by low signal-to-noise ratio, which hinders reliable identification of jumps. At high measurement rates, measurement of the anti-symmetric mode inevitably dephases the symmetric mode, due to its dispersive coupling to the readout resonator ($2\chi_s/2\pi = 0.84\,\mathrm{MHz}$, almost equal to the dispersive shift of the anti-symmetric mode; \hyperref[app:chi_sm]{Appendix~I}). We attribute the decrease in efficiency to this dephasing, which suppresses the coherent component of photon absorption and hence the absorption efficiency.

We describe this trade-off between signal-to-noise ratio and measurement-induced dephasing of the symmetric mode with a simple model, resulting in the expression:
\begin{equation}
\eta_\mathrm{cont}(\Gamma_\mathrm{m}) = F_\mathrm{ro}^\mathrm{cont}(\Gamma_\mathrm{m})\, \eta_\mathrm{abs}(\Gamma_\mathrm{m}),
\end{equation}
where $F_\mathrm{ro}^\mathrm{cont}$ is the continuous readout fidelity and $\eta_\mathrm{abs}$ is the absorption efficiency obtained by including dephasing components in the non-Hermitian Hamiltonian (\hyperref[app:theory]{Appendix~E}). The model agrees well with the experimental data without any fit parameters.

\section{III. DISCUSSION}

In summary, we demonstrate a continuously operated detector of microwave photons with an efficiency of $\eta_\mathrm{cont} = 0.47$, a dark count rate of $\Gamma_\mathrm{dark}^\mathrm{cont} = 1.3\,\mathrm{kHz}$, a temporal resolution of $1\,\mu\mathrm{s}$, and a dead time of $15\,\mu\mathrm{s}$. Our device, an artificial molecule formed by two coupled transmons, transfers photons incident on its symmetric mode to a continuously monitored anti-symmetric mode via a driven-dissipative process. To the best of our knowledge, this work constitutes the first experimental demonstration of a continuously operated microwave photodetector that can resolve single photons. At present, the efficiency is limited by spurious measurement backaction, which suppresses the coherent component of the photon-transfer process. This limitation is not intrinsic to the protocol, but instead arises from the present device architecture, in which measurement of the anti-symmetric mode also dephases other modes of the artificial molecule. A route to improved performance is to engineer a four-level system with mode-selective readout. Our theoretical model predicts that, in such a design, the photon-capture efficiency would be insensitive to dephasing of the monitored mode. This would also enable faster readout, which, coupled with a reset protocol, could reduce the dead time below $1\,\mu\mathrm{s}$. The ability to detect microwave photons continuously is particularly valuable in experiments where photon-emission events are stochastic, such as fluorescence from a single spin~\cite{wang2023} or emission from a cavity haloscope in an axion search experiment~\cite{braggio2025, dixit2021}. Our work therefore establishes always-on photodetection as a practical tool for microwave quantum optics.

\section*{ACKNOWLEDGMENTS}

We thank Jiaying Yang and Ludvig Nordqvist for helpful discussions on the analysis of continuous readout signals. We thank the VTT Technical Research Center of Finland for providing the TWPA used in this experiment. The device was fabricated at Myfab Chalmers and its design was assisted by the Python package QuCAT~\cite{gely2020}. We acknowledge financial support from the Swedish Research Council, the Knut and Alice Wallenberg Foundation through the Wallenberg Center for Quantum Technology (WACQT), from the European Research Council via Grant No. 101041744 ESQuAT and from the European Union via Grant No. 101080167 ASPECTS.

\bibliographystyle{apsrev4-2}
\bibliography{References}
\clearpage
\appendix
\onecolumngrid

\raggedbottom

\section{APPENDIX A: DEVICE FABRICATION AND MEASUREMENT SETUP}\label{app:wiring}

\begin{figure}[b]
    \centering
    \includegraphics[width=0.92\textwidth]{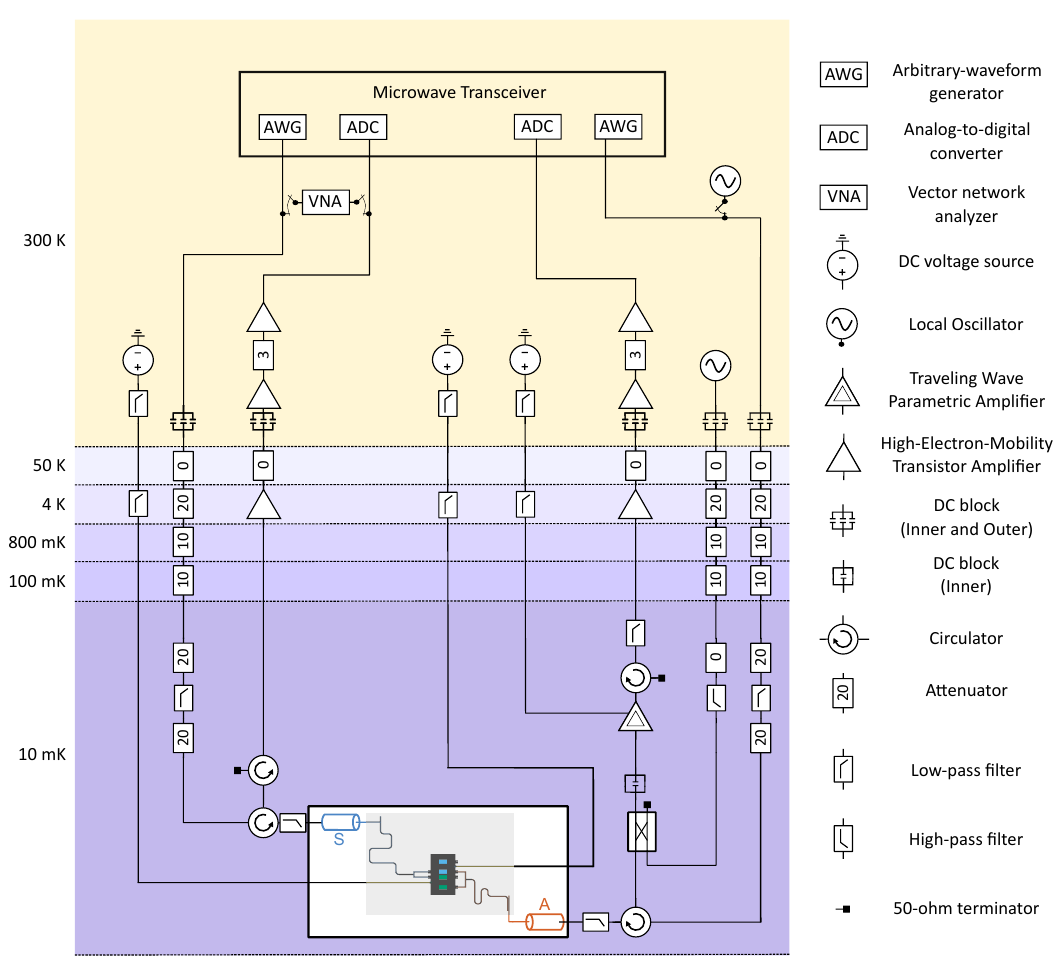}
    \caption{Complete experimental setup. See text for description.}
    \label{fig:wiring}
\end{figure}

The device is fabricated on a high-resistivity two-inch silicon wafer. Capacitors and transmission lines are patterned on a $300\,\mathrm{nm}$ thick aluminum ground plane. Josephson junctions and patches are defined with a Raith EBPG5200 electron-beam lithography system and deposited in a Plassys MEB550S electron-beam evaporator. The junctions are deposited by double-angle shadow evaporation. Airbridges are written and deposited on top to connect isolated ground planes.

\hyperref[fig:wiring]{Figure~\ref*{fig:wiring}} shows a schematic of the experimental setup used to study the device. The device is packaged in a copper sample holder and wire-bonded to a printed circuit board using aluminum bond wires. The sample holder is mounted on the mixing-chamber stage of a dilution refrigerator and enclosed within a copper can and a \textmu-metal shield. Microwave tones are routed to the device through heavily attenuated coaxial lines. Both waveguides, A and S, are measured in a reflection configuration. The outgoing field from waveguide~A is amplified first by a traveling-wave parametric amplifier mounted at the mixing-chamber stage and subsequently by a high-electron-mobility transistor (HEMT) amplifier at the $4\,\mathrm{K}$ stage, whereas the outgoing field from waveguide~S is amplified by a HEMT amplifier at the $4\,\mathrm{K}$ stage only. We use the microwave transceiver Presto (Intermodulation Products AB) to generate and acquire microwave signals.

\section{APPENDIX B: SYMMETRIC MODE SPECTROSCOPY AND PROBE POWER CALIBRATION}\label{app:sm_fit}
The reflection coefficient $r$ for a two-level system coupled to the end of a waveguide can be derived from the Lindblad master equation and input-output theory~\cite{scigliuzzo2020, lu2021b}. In our device, we treat the symmetric mode as a two-level system. Neglecting the thermal occupation, we model the reflection coefficient for the symmetric mode, as shown in \hyperref[fig:fig1]{Fig.~\ref*{fig:fig1}c,d}, by the equation:

\begin{equation}
r(\omega-\omega_s)
= 1-\frac{i\Gamma_s\Gamma_{1s}\left(\omega-\omega_s-i\Gamma_{2s}\right)}
{\Omega_\mathrm{pr}^2\Gamma_{2s}+\Gamma_{1s}\left[(\omega-\omega_s)^2+\Gamma_{2s}^2\right]}
\end{equation}

Here, $\omega$ is the probe frequency, $\Gamma_s$ denotes the coupling rate of the symmetric mode to waveguide~S, $\Gamma'_s$ the decay rate into all other channels, and $\Gamma_{s\phi}$ the pure dephasing rate of the symmetric mode, with $\Gamma_{1s}=\Gamma_s+\Gamma'_s$ and $\Gamma_{2s}=(\Gamma_s+\Gamma'_s)/2+\Gamma_{s\phi}$. $\Omega_\mathrm{pr}$ is the Rabi rate induced by the probe field. A global fit of this expression to the measured data yields $\Gamma_s$ and $\Omega_\mathrm{pr}$, from which we determine the power at the device, $P_\mathrm{device}$, as~\cite{scigliuzzo2020}:

\begin{equation}
 P_\mathrm{device} = \frac{\hbar \omega_s \Omega_\mathrm{pr}^2}{4\Gamma_s }
\end{equation}

Comparing $P_\mathrm{device}$ with the applied power gives the total attenuation of the input line to waveguide~S. We then use this attenuation to determine the photon flux in continuous-detection measurements and the mean photon number of the Gaussian photon pulses used in cyclic detection.

\section{APPENDIX C: SINGLE-SHOT READOUT}\label{app:singleshot}
\begin{figure}[H]
    \centering
    \includegraphics[width=3in]{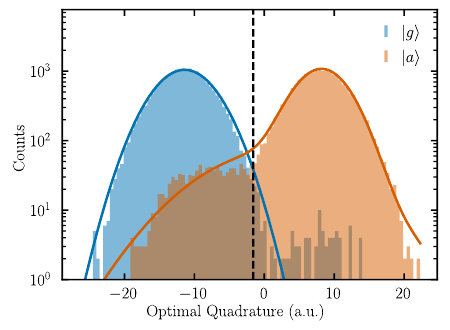}
    \caption{Histograms of single-shot readout statistics, with the molecule prepared in $\ket{g}$ (blue) and $\ket{a}$ (orange). Counts are plotted against the optimal readout quadrature. Solid lines show the fitted Gaussian mixture models and the dashed black line indicates the decision boundary.}
    \label{fig:singleshot}
\end{figure}

We obtain single-shot statistics by preparing the molecule in $\ket{g}$ and $\ket{a}$, the latter using a $\pi$ pulse. Each histogram comprises $20000$ single shots acquired with a readout pulse of duration $1\,\mu\mathrm{s}$. We rotate the data in the IQ plane so that the separation between the two state distributions lies entirely along one quadrature, and retain only that quadrature. We fit a Gaussian mixture model to the single-shot data for each preparation, using one component for $\ket{g}$ and two for $\ket{a}$, the second accounting for shots in which the molecule decays to $\ket{g}$ during the readout pulse. We take the decision boundary to be the midpoint between the means of the largest Gaussian component of each histogram. We find a pulsed readout fidelity of $F_\mathrm{ro}^\mathrm{cyc} = 1 - P(g\vert a) - P(a\vert g) = 0.9474$. The separation error $\epsilon_\mathrm{sep}$, the overlap of the two Gaussian distributions at the decision boundary, is $0.0015$, and the transition errors, which arise from state changes during the readout pulse, are $\epsilon_{\mathrm{g}\rightarrow\mathrm{a}} = 0.005$ and $\epsilon_{\mathrm{a}\rightarrow\mathrm{g}} = 0.0476$~\cite{walter2017}.

\section{APPENDIX D: CHARACTERIZING THE PUMP AND THE \texorpdfstring{$\ket{2-}$}{|2-⟩} STATE}\label{app:pump}
\begin{figure}[H]
    \centering
    \includegraphics[width=\textwidth]{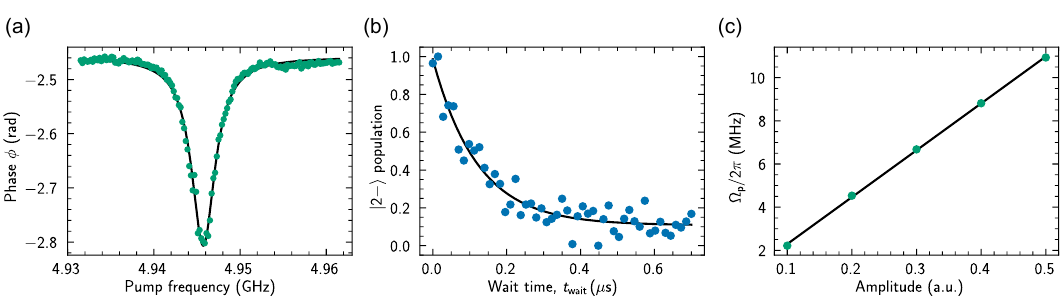}
    \caption{
    \textbf{a.} Pump frequency calibration. The phase of the readout resonator response is plotted against the pump-tone frequency while the symmetric mode is simultaneously saturated. The solid line shows a Lorentzian fit to the data.
    \textbf{b.} Decay rate calibration. The population of $\ket{2-}$ is plotted as a function of the wait time $t_\mathrm{wait}$ between the $\pi$ pulses and the population measurement. The solid line shows an exponential fit to the data.
    \textbf{c.} Pump amplitude calibration using the Autler-Townes splitting. The pump Rabi rate $\Omega_\mathrm{p}$, obtained as half the measured frequency splitting, is plotted against the pump amplitude set at room temperature, in arbitrary units. The solid line shows a linear fit to the data.
    }
    \label{fig:twominus}
\end{figure}
We first calibrate the frequency of the pump. We drive the symmetric mode weakly with a continuous coherent tone while sweeping the pump tone frequency at fixed pump power, and simultaneously probe the readout resonator weakly with a continuous readout tone. Excitations in the symmetric mode are transferred to the anti-symmetric mode through the $\ket{2-}$ state. The resulting steady-state population of $\ket{a}$ shifts the readout resonator, so the resonator phase is maximally displaced when the pump is resonant with the $\ket{s}\leftrightarrow\ket{2-}$ transition. We plot the phase of the resonator response in \hyperref[fig:twominus]{Fig.~\ref*{fig:twominus}a} against the frequency of the pump tone. The $\ket{s}\leftrightarrow\ket{2-}$ transition frequency is extracted from a Lorentzian fit to the data, giving $\omega_{s,2-}/2\pi = 4.946\,\mathrm{GHz}$. 

Next, we extract the rate $\Gamma_{2-}$ at which $\ket{2-}$ decays into $\ket{a}$. We prepare $\ket{2-}$ with two successive $\pi$ pulses, on the $\ket{g}\leftrightarrow\ket{a}$ and $\ket{a}\leftrightarrow\ket{2-}$ transitions. We then measure the population of $\ket{a}$ after a variable wait time $t_\mathrm{wait}$. \hyperref[fig:twominus]{Figure~\ref*{fig:twominus}b} plots the population of $\ket{2-}$ against $t_\mathrm{wait}$. An exponential fit to the data yields $\Gamma_{2-}/2\pi = 1.32\,\mathrm{MHz}$.

Finally, we calibrate the Rabi rate $\Omega_\mathrm{p}$ of the pump using the Autler-Townes splitting of the $\ket{s}\leftrightarrow\ket{2-}$ transition. We perform reflection spectroscopy of the $\ket{g}\leftrightarrow\ket{s}$ transition while sweeping the amplitude of the pump tone. The measured frequency splitting equals $2\Omega_\mathrm{p}$, and we plot the resulting Rabi rate against the pump amplitude set at room temperature (\hyperref[fig:twominus]{Fig.~\ref*{fig:twominus}c}). From the slope of a linear fit, the pump amplitude used in the photodetection experiments corresponds to $\Omega_\mathrm{p}/2\pi = 2.17\,\mathrm{MHz}$.

\section{APPENDIX E: MODEL FOR THE CONTINUOUS DETECTION EFFICIENCY}\label{app:theory}

We model the continuous detection efficiency as the product of the probability $\eta_\mathrm{abs}$ that an incident photon is absorbed, and the fidelity $F_\mathrm{ro}^\mathrm{cont}$ with which that excitation is subsequently identified in the continuous readout record,
\begin{equation}
    \eta_\mathrm{cont} = F_\mathrm{ro}^\mathrm{cont}\,\eta_\mathrm{abs}.
\end{equation}

Both factors are set by the strength of the continuous readout, but they do not respond to it through the same quantity. The first quantity is the measurement-induced dephasing rate $\Gamma_\phi$, the rate at which the readout suppresses the coherence of the monitored transition and, equivalently, accumulates information about its state. The second is the readout-induced broadening $w_\phi$, the increase in the half width at half maximum of that transition. The two are proportional only as long as the line stays Lorentzian, which over the range of readout amplitudes swept in \hyperref[fig:fig3]{Fig.~\ref*{fig:fig3}d} it does not. Equivalently, the coherence of the monitored transition decays as~\cite{gambetta2006}
\begin{equation}
    C(t) = e^{-\beta_m(t)},\qquad
    \beta_m(t) = 4\bar{n}\,\theta_0^2\left(\frac{\kappa_r t}{2}-1+e^{-\kappa_r t/2}\right),
    \label{eq:betam}
\end{equation}
where $\theta_0 = \arctan\left(2\chi_a/\kappa_r\right)$. At timescales greater than $1/\kappa_r$, the decay is exponential with a decay rate $\Gamma_\phi\simeq2\bar{n}\theta_0^2\kappa_r$. At timescales smaller than $1/\kappa_r$, the decay is slower, and is well approximated by the increase in linewidth of the qubit spectrum, $w_\phi$. Absorption occurs on the shorter timescale, $1/\kappa_D<1/\kappa_r$, and readout on the longer one, $t_\mathrm{int}\gg1/\kappa_r$, so $\eta_\mathrm{abs}$ is set by $w_\phi$ and $F_\mathrm{ro}^\mathrm{cont}$ by $\Gamma_\phi$.

\subsection{1. PHOTON CAPTURE}
The capture process involves the symmetric mode $\ket{s}$, which is coupled to waveguide~S at rate $\Gamma_s$, and the doubly excited state $\ket{2-}$, which decays into $\ket{a}$ at rate $\Gamma_{2-}$. The pump drives the $\ket{s}\leftrightarrow\ket{2-}$ transition at Rabi rate $\Omega_\mathrm{p}$. We describe the dynamics with the effective non-Hermitian Hamiltonian in the basis $\{\ket{s},\ket{2-}\}$ ($\hbar=1$):
\begin{equation}
    H_\mathrm{eff} =
    \begin{pmatrix}
        -i\left(\dfrac{\Gamma_s}{2}+\dfrac{w_\phi}{4}\right) & \Omega_\mathrm{p} \\[2ex]
        \Omega_\mathrm{p} & -i\left(\dfrac{\Gamma_{2-}}{2}+\dfrac{w_\phi}{2}\right)
    \end{pmatrix}.
\end{equation}
Here $w_\phi$ is the readout-induced broadening introduced above, which we take from the photon shot-noise line shape evaluated at the calibrated photon number (\hyperref[app:measurement_rate]{Appendix~G}). Continuous monitoring of the anti-symmetric mode also broadens the remaining modes of the artificial molecule. Taking the broadening of a state to scale with its excitation number, the singly excited state $\ket{s}$ acquires an excess width $w_\phi/2$ and the doubly excited state $\ket{2-}$ an excess width $w_\phi$; the associated coherences decay at half these rates, which is what enters $H_\mathrm{eff}$.

For a single photon incident on the symmetric mode at detuning $\delta$ from resonance, the amplitude for conversion into an excitation of $\ket{a}$ follows from the resolvent $G(\delta)=\left(H_\mathrm{eff}-\delta\,\mathbf{1}\right)^{-1}$ as $t(\delta) = -i\sqrt{\Gamma_s\Gamma_{2-}}\,G_{12}(\delta)$, so that $\eta_\mathrm{abs}=\abs{t(\delta)}^2$. Evaluating the inverse gives
\begin{equation}
    \eta_\mathrm{abs}(\delta) = \frac{64\,\Gamma_s\Gamma_{2-}\Omega_\mathrm{p}^2}
    {4\delta^2\left(2\Gamma_s+2\Gamma_{2-}+3w_\phi\right)^2
    +\left[\left(2\Gamma_s+w_\phi\right)\left(\Gamma_{2-}+w_\phi\right)+8\Omega_\mathrm{p}^2-8\delta^2\right]^2}.
\end{equation}
For a resonant photon, $\delta=0$, this reduces to
\begin{equation}
    \eta_\mathrm{abs} = \frac{64\,\Gamma_s\Gamma_{2-}\Omega_\mathrm{p}^2}
    {\left[\left(2\Gamma_s+w_\phi\right)\left(\Gamma_{2-}+w_\phi\right)+8\Omega_\mathrm{p}^2\right]^2}.
\end{equation}

\begin{table}[H]
    \centering
    \caption{Parameters used in the model. All rates are quoted as ordinary frequencies, $X/2\pi$.}
    \label{tab:theory_params}
    \begin{ruledtabular}
    \begin{tabular}{lll}
    Parameter & Value & Calibration \\
    \hline
    $\Gamma_s/2\pi$ & $7.26\,\mathrm{MHz}$ & Coupling of $\ket{s}$ to waveguide~S (\hyperref[app:sm_fit]{Appendix~B}) \\
    $\Gamma_{2-}/2\pi$ & $1.32\,\mathrm{MHz}$ & Decay rate of $\ket{2-}$ into $\ket{a}$ (\hyperref[app:pump]{Appendix~D}) \\
    $\Omega_\mathrm{p}/2\pi$ & $2.17\,\mathrm{MHz}$ & Pump Rabi rate from Autler-Townes splitting (\hyperref[app:pump]{Appendix~D}) \\
    $\chi_a/2\pi$ & $0.434\,\mathrm{MHz}$ & Dispersive shift per photon, sets $\Gamma_\phi$ and $w_\phi$ (\hyperref[app:measurement_rate]{Appendix~G}) \\
    $\kappa_r/2\pi$ & $1.8\,\mathrm{MHz}$ & Readout resonator linewidth, sets $\Gamma_\phi$ and $w_\phi$ \\
    $\eta$ & $0.11$ & Measurement efficiency (\hyperref[app:measurement_rate]{Appendix~G}) \\
    $t_\mathrm{int}$ & $5\,\mu\mathrm{s}$ & Matched-filter integration time \\
    \end{tabular}
    \end{ruledtabular}
\end{table}

\subsection{2. READOUT FIDELITY}
Jumps are identified from the measurement record after matched filtering, with an integration time $t_\mathrm{int}=5\,\mu\mathrm{s}$ set by the five-tap kernel applied to the $1\,\mu\mathrm{s}$-binned record. Over that window the readout accumulates a signal-to-noise ratio $\sqrt{\Gamma_\mathrm{m}t_\mathrm{int}}=\sqrt{\eta\,\Gamma_\phi t_\mathrm{int}}$, where $\eta$ is the measurement efficiency. A jump arrives at a random time within the window and only the part of the window that follows it carries signal, so half of the window contributes on average, and we take the readout fidelity to be
\begin{equation}
    F_\mathrm{ro}^\mathrm{cont} = 1-\mathrm{erfc}\left(\tfrac{1}{2}\sqrt{\eta\,\Gamma_\phi t_\mathrm{int}}\right).
\end{equation}

The two factors in $\eta_\mathrm{cont}$ therefore depend on the readout strength in opposite ways: increasing the readout power improves $F_\mathrm{ro}^\mathrm{cont}$ but suppresses $\eta_\mathrm{abs}$, which produces the maximum observed in \hyperref[fig:fig3]{Fig.~\ref*{fig:fig3}d}.

All parameters entering the model are calibrated independently, as summarized in \hyperref[tab:theory_params]{Table~\ref*{tab:theory_params}}, so that the solid line in \hyperref[fig:fig3]{Fig.~\ref*{fig:fig3}d} contains no free parameters. The model reproduces the maximum we observe, predicting an efficiency of $0.51$ at $\Gamma_\mathrm{m}/2\pi = 0.11\,\mathrm{MHz}$ against a measured $0.47$ at $0.10\,\mathrm{MHz}$. It does not account for the finite lifetime of $\ket{a}$.

\section{APPENDIX F: DETECTION BANDWIDTH}\label{app:bandwidth}
\begin{figure}[H]
    \centering
    \includegraphics[width=3in]{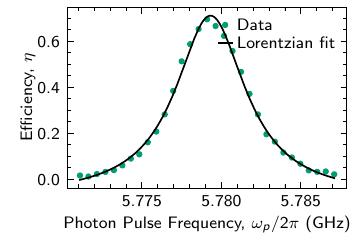}
    \caption{Cyclic detection efficiency $\eta_\mathrm{cyc}$ as a function of the photon pulse frequency $\omega_\mathrm{ph}/2\pi$. The solid line shows a Lorentzian fit to the data, whose full width at half maximum gives the detection bandwidth $\kappa_D/2\pi$.}
    \label{fig:bandwidth}
\end{figure}

We measure the detection bandwidth by repeating the cyclic photodetection sequence while sweeping the frequency $\omega_\mathrm{ph}$ of the $500\,\mathrm{ns}$ photon pulse across the symmetric mode. We plot the cyclic detection efficiency $\eta_\mathrm{cyc}$ against $\omega_\mathrm{ph}/2\pi$ in \hyperref[fig:bandwidth]{Fig.~\ref*{fig:bandwidth}}, together with a Lorentzian fit. The response is a single Lorentzian centered at $\omega_0/2\pi = 5.7793\,\mathrm{GHz}$, coinciding with the frequency of the symmetric mode, with a full width at half maximum of $\kappa_D/2\pi = 5.37\,\mathrm{MHz}$, which we take as the detection bandwidth. 

\section{APPENDIX G: CALIBRATION OF THE MEASUREMENT RATE}\label{app:measurement_rate}
\begin{figure}[H]
    \centering
    \includegraphics[width=\textwidth]{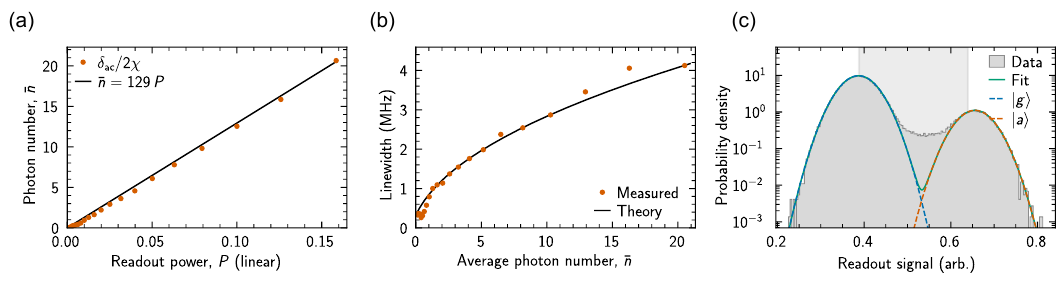}
    \caption{Calibration of the measurement rate.
    \textbf{a.} Average photon number $\bar{n}$ in the readout resonator against the readout power $P$ applied to it, in linear units. Solid line is a linear fit.
    \textbf{b.} Half width at half maximum of the $\ket{g}\leftrightarrow\ket{a}$ line against $\bar{n}$. The solid line is the photon shot-noise line shape of Ref.~\cite{schuster2007}, evaluated with $\chi_a$ and $\kappa_r$ fixed at their independently measured values.
    \textbf{c.} Histogram of a $1\,\mathrm{s}$ continuous readout record integrated in $5\,\mu\mathrm{s}$ bins, at the readout power corresponding to $\bar{n} = 2.6$. The solid line is a two-component Gaussian mixture fitted to the record; dashed lines are its $\ket{g}$ and $\ket{a}$ components. The shaded band marks the samples excluded from the fit.
    }
    \label{fig:starkshift}
\end{figure}

The measurement rate is given by $\Gamma_\mathrm{m} = \eta\,\Gamma_\phi$ where $\Gamma_\phi$ is the measurement-induced dephasing rate of the anti-symmetric mode and $\eta$ is the measurement efficiency of the amplification chain. We first calibrate $\bar{n}$ against the readout power set at room temperature with an ac-Stark shift measurement: we perform spectroscopy of the $\ket{g}\leftrightarrow\ket{a}$ transition while applying a readout tone of variable power $P$, and track the frequency shift $\delta_\mathrm{ac}$~\cite{schuster2007} (\hyperref[fig:starkshift]{Fig.~\ref*{fig:starkshift}a}). Each photon in the resonator shifts the transition by $2\chi_a$ and the resonator responds linearly to the drive, so $\delta_\mathrm{ac} = 2\chi_a\bar{n}$ with $\bar{n} = kP$. 
The same spectra give the linewidth of the transition (\hyperref[fig:starkshift]{Fig.~\ref*{fig:starkshift}b}). Photon-number fluctuations in the resonator translate into frequency fluctuations of the anti-symmetric mode and broaden its half width at half maximum from $0.33\,\mathrm{MHz}$ at $\bar{n}\rightarrow0$ to $4.1\,\mathrm{MHz}$ at $\bar{n} = 20.5$, sublinearly in $\bar{n}$: the line is not Lorentzian in this regime. The solid line is the photon shot-noise line shape of Ref.~\cite{schuster2007}, which is the Fourier transform of the coherence decay $C(t) = e^{-\beta_m(t)}$ of Eq.~(\ref{eq:betam}); we evaluate it with $\chi_a$ and $\kappa_r$ fixed at their measured values and with the $\bar{n}\rightarrow0$ width as its only other input, so that no parameter is fitted to these data. From the same expression we obtain the two quantities that \hyperref[app:theory]{Appendix~E} needs: the readout-induced broadening $w_\phi(\bar{n})$, taken as the excess width over the $\bar{n}\rightarrow0$ value, and the dephasing rate $\Gamma_\phi = \beta_m(t_\mathrm{int})/t_\mathrm{int}$, over $t_\mathrm{int} = 5\,\mu\mathrm{s}$.

We extract the measurement efficiency $\eta$ from the continuous detection records themselves. At each readout power we integrate one second of readout signal in $5\,\mu\mathrm{s}$ bins, histogram the result, and fit a two-component Gaussian mixture to it, from which we take $\mathrm{SNR} = (\mu_a-\mu_g)/\sigma_g$ (\hyperref[fig:starkshift]{Fig.~\ref*{fig:starkshift}c}); we exclude the interval between the two means from the fit, to prevent the decay of $\ket{a}$ from biasing the means. The measurement efficiency is then~\cite{bultink2018, clerk2010a}
\begin{equation}
    \eta = \frac{\mathrm{SNR}^2}{8\beta_m},
\end{equation}
with $\beta_m$ evaluated at the same $\bar{n}$ and $t_\mathrm{int}$; in this convention, $\eta = 0.5$ is the quantum limit for phase-preserving amplification. Across the seven readout powers the SNR runs from $2.5$ to $19.6$ and $\eta$ from $0.10$ to $0.12$, with a mean of $\eta = 0.11$.
\section{APPENDIX H: DEAD TIME}\label{app:deadtime}
\begin{figure}[H]
    \centering
    \includegraphics[width=3in]{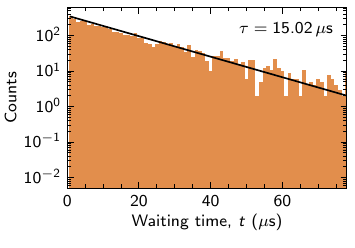}
    \caption{Distribution of the times the detector spends in $\ket{a}$ during continuous operation, obtained from a $1\,\mathrm{s}$ measurement record at an incident photon flux of $\Phi_\mathrm{ph} = 10^4\,\mathrm{photons/s}$. The solid line is an exponential fit, whose decay constant $\tau$ we take as the dead time.}
    \label{fig:deadtime}
\end{figure}

Once a captured photon has excited the anti-symmetric mode, the detector cannot register a further click until $\ket{a}$ has relaxed back to $\ket{g}$. The duration of this blind interval is the dead time $t_\mathrm{dead}$, which we extract from the dwell time distribution of the $\ket{a}$ state.

We analyze a single $1\,\mathrm{s}$ measurement record, integrated in $1\,\mu\mathrm{s}$ bins as in the main text, taken at $\Phi_\mathrm{ph} = 10^4\,\mathrm{photons/s}$ and a measurement rate of $\Gamma_\mathrm{m}/2\pi = 0.8\,\mathrm{MHz}$. We assign each sample to $\ket{g}$ or $\ket{a}$ by fitting a two-component Gaussian mixture to the rotated record and identifying the component of larger weight with $\ket{g}$. This assignment is independent of the matched-filter jump counting used in the main text. We histogram the dwell time in state $\ket{a}$ and fit it to an exponential distribution. We interpret the time constant of this distribution as the dead time of the detector, and find $t_\mathrm{dead} = 15\,\mu\mathrm{s}$, which is close to the measured relaxation time of $\ket{a}$ ($T_1 = 18\,\mu\mathrm{s}$).

\section{APPENDIX I: DISPERSIVE COUPLING OF SYMMETRIC MODE AND READOUT RESONATOR}\label{app:chi_sm}
\begin{figure}[H]
    \centering
    \includegraphics[width=3in]{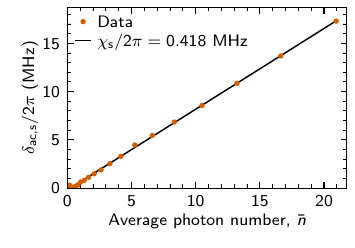}
    \caption{ac-Stark shift $\delta_\mathrm{ac,s}/2\pi$ of the $\ket{g}\leftrightarrow\ket{s}$ transition against the average photon number $\bar{n}$ in the readout resonator, calibrated as in \hyperref[app:measurement_rate]{Appendix~G}. The solid line is a linear fit, whose slope gives $2\chi_s$.}
    \label{fig:starkshiftSM}
\end{figure}

The readout resonator is dispersively coupled to the symmetric mode as well as to the anti-symmetric mode. We calibrate the readout power in terms of the average photon number $\bar{n}$ in the resonator using the measurement of \hyperref[app:measurement_rate]{Appendix~G}. We then perform an ac-Stark shift measurement on the symmetric mode and plot the shift against $\bar{n}$ (\hyperref[fig:starkshiftSM]{Fig.~\ref*{fig:starkshiftSM}}), whose slope gives $2\chi_s/2\pi = 0.837\,\mathrm{MHz}$.

Consequently, the continuous readout tone that monitors $\ket{a}$ also dephases $\ket{s}$, broadening it by an amount comparable to the broadening of the anti-symmetric mode. This justifies the readout-induced broadening of $\ket{s}$, and of $\ket{2-}$, included in the capture model of \hyperref[app:theory]{Appendix~E}, through which the measurement backaction of the continuous readout suppresses the absorption efficiency.

\end{document}